\documentclass[useAMS,usenatbib]{mn2e}
\usepackage{mathptmx}
\usepackage{amssymb}
\usepackage{aecompl}
\usepackage[T1]{fontenc}
\usepackage{afterpage}

\usepackage{graphicx}
\usepackage[update,prepend]{epstopdf}

\newcommand{\Msun}{\ensuremath{{\rm M}_{\odot}}}

\title[Molecular opacities and AGB evolution]{The effect of including molecular opacities of variable composition on the evolution of intermediate-mass AGB stars}
\author[Fishlock et al.]{C. K. Fishlock$^{1}$\thanks{cherief@mso.anu.edu.au}, A. I. Karakas$^{1}$ and R. J.  Stancliffe$^{1,2}$\\
$^{1}$Research School of Astronomy \& Astrophysics, Australian National University,
Canberra ACT 2611, Australia\\
$^{2}$Argelander-Institut f\"{u}r Astronomie, Auf dem H\"{u}gel 71, D-53121 Bonn, Germany}

\begin{document}

\date{Accepted 2013 November 29. Received 2013 November 28; in original form 2013 April 3}

\pagerange{\pageref{firstpage} - \pageref{lastpage}} \pubyear{2013}

\maketitle

\label{firstpage}

\begin{abstract}

Calculations from stellar evolutionary models of low- and intermediate-mass asymptotic giant branch (AGB) stars provide predictions of elemental abundances and yields for comparison to observations.  However, there are many uncertainties that reduce the accuracy of these predictions. One such uncertainty involves the treatment of low-temperature molecular opacities that account for the surface abundance variations of C, N, and O. A number of prior calculations of intermediate-mass AGB stellar models that incorporate both efficient third dredge-up and hot bottom burning  include a molecular opacity treatment which does not consider the depletion of C and O due to hot bottom burning. 
Here we update the molecular opacity treatment and investigate the effect of this improvement on calculations of intermediate-mass AGB stellar models.
We perform tests on two masses, 5~$\Msun$ and 6~$\Msun$, and two metallicities, $Z~=~0.001$ and $Z~=~0.02$, to quantify the variations between two opacity treatments. We find that several evolutionary properties (e.g. radius, $T_{\rm eff}$ and $T_{\rm bce}$) are dependent on the opacity treatment. Larger structural differences occur for the $Z~=~0.001$ models compared to the $Z~=~0.02$ models indicating that the opacity treatment has a more significant effect at lower metallicity. As a consequence of the structural changes, the predictions of isotopic yields are slightly affected with most isotopes experiencing changes up to 60 per cent for the $Z~=~0.001$ models and 20 per cent for the $Z~=~0.02$ models. Despite this moderate effect, we conclude that it is more fitting to use variable molecular opacities for models undergoing hot bottom burning. 

\end{abstract}


\section{Introduction}

The asymptotic giant branch (AGB) phase is the last nuclear burning stage for low- and intermediate-mass stars with initial masses of between approximately 0.8~$\Msun$ to $\sim 8$~$\Msun$ \citep[see][for a review]{Herwig05}. During their evolution on the AGB, stars in this mass range experience mixing episodes that dredge up newly synthesised material which contains the products of He-burning and neutron-capture nucleosynthesis into the convective envelope \citep[see][for a review]{Busso99}. Combined with strong mass loss that ejects the stellar envelope into the interstellar medium, low- and intermediate-mass AGB stars are important contributors to the chemical evolution of the Galaxy \citep[for example,][]{Romano10,Kobayashi11,RecioBlanco12}. This study is concerned with the evolution of intermediate-mass stars on the AGB with an initial mass $\gtrsim$ 4~$\Msun$. These stars produce substantial primary nitrogen and are predicted to produce nitrogen enhanced metal poor stars in the early Galaxy \citep[e.g.][]{Pols12}. Intermediate-mass AGB stars are also implicated in the chemical evolution of globular clusters \citep{Gratton04,Ventura11}. However, the modelling of the evolution and nucleosynthesis of low- and intermediate-mass AGB stars is dependent on highly uncertain input physics including convection, mass loss, reaction rates and opacity.

Briefly, the structure of an AGB star consists of an electron-degenerate CO core surrounded by a He-burning shell and a H-burning shell which are separated by the He-intershell. This is encompassed by a deep convective envelope. During the thermally-pulsing AGB (TP-AGB) phase, the star undergoes periodic thermal pulses (TPs) due to instabilities in the He-burning shell \citep{Herwig05,Busso99}. During the evolution of a TP-AGB star, two known mechanisms can occur which can  periodically alter the surface composition: third dredge-up (TDU) and hot bottom burning (HBB).  When the He-burning shell becomes unstable it drives an expansion of the star in order to release the energy generated by He-burning. A flash driven convective zone forms in the He-intershell mixing material, mainly $^{12}$C, from the base of the He-shell throughout the intershell region. The outer layers have been cooled by the expansion and the H-burning shell has been almost extinguished which allows the outer convective envelope to move inwards. If the convective envelope reaches into the He-intershell, $^{12}$C and the products of slow neutron-capture nucleosynthesis (the $s$-process) are mixed to the surface. This process of TDU can occur multiple times and is responsible for creating carbon-rich stars with a C/O ratio greater than unity.  

HBB occurs in stars with masses greater than approximately 4~$\Msun$ depending on metallicity and the input physics used. HBB takes place during the interpulse phase between TPs when the base of the outer convective envelope penetrates into the H-burning shell and becomes hot enough to sustain proton-capture nucleosynthesis. The CNO cycle converts $^{12}$C and $^{16}$O that has been mixed into the envelope by TDU into predominately $^{14}$N.  Therefore HBB can prevent the stellar surface from becoming carbon rich  by decreasing the $^{12}$C abundance in the envelope. \citet{Frost98} presented evolutionary calculations of intermediate-mass AGB stellar models and found that the surface C/O ratio initially decreases due to HBB. However, mass loss slowly erodes the envelope causing a decrease in envelope mass. This results in a lower temperature at the base of the convective envelope which causes HBB to cease. There is also less dilution of the dredged-up material. This means that the C/O ratio starts to increase at the end of the TP-AGB phase, and in some cases to reach above unity.  

Mass loss and convection have been shown to dominate modelling uncertainties in intermediate-mass AGB models \citep{Ventura05a, Ventura05b, Stancliffe07,Karakas12}. However, other uncertainties including reaction rates and opacities have been shown to affect the stellar structure and therefore the yields \citep{Izzard07,Ventura09}. In recent years there has been considerable work developing accurate low-temperature molecular opacities for stellar evolution calculations \citep{Marigo09}. Due to these improvements to the opacity input physics it is now possible to quantify the effect of the updated opacities on the stellar evolution calculations and yield predictions \citep{Ventura09,Weiss09}. 

Intermediate-mass AGB stars have been shown to have effective temperatures cool enough for dust and molecule formation at solar metallicity and in the Magellanic Clouds \citep{GarciaHernandez09}. In particular, we also show in this study that low-metallicity AGB stars become cool enough to form molecules particularly once they become carbon rich.  
 The opacity tables of \citet{Alexander94}, and later \citet{Ferguson05}, include a detailed treatment of the inclusion of molecules to the total opacity at low temperatures where $T \lesssim 10^4$ K. These tables, however, are only available for solar or scaled-solar composition. As previously mentioned, low- and intermediate-mass AGB stars undergo mixing episodes that alter their surface composition in a complex way. In particular, low-mass AGB stars can become carbon rich and intermediate-mass stars can display a range of behaviours for the C/O ratio.  \citet{Marigo02} shows that, at the transition point when the C/O ratio goes from below unity to above unity, the dominant source of molecular opacity changes from oxygen-bearing molecules to carbon-bearing molecules. In AGB models, this causes a sudden decrease of the effective temperature and an expansion in radius which in turn increases the rate of mass loss. It is therefore necessary to use low-temperature molecular opacities that follow the change in the C/O ratio with time. The low-temperature opacity tables of \citet{Lederer09} only account for an enhancement in C and N compared to initial abundances whereas the \AE SOPUS opacity tables of \citet{Marigo09} are able to account for the depletion and enhancement of C, N and the C/O ratio.

Using the \AE SOPUS molecular opacity tables, the effects of molecular opacities on the evolution of AGB stellar models  have been investigated by \citet{Ventura09} and \citet{Ventura10}. Using two different opacity treatments they determine that the yields of C, N, O and Na can be significantly altered depending on the opacity prescription used. \citet{Ventura10} stress that it is important to use a molecular opacity treatment that accounts for variations in the surface CNO abundance when the C/O ratio exceeds unity. They find a minimum threshold mass for AGB stellar models ($\geq$ 3.5~$\Msun$ for $Z~=~0.001$) where the use of opacity tables that account for the variations in the surface CNO abundance becomes less important. This is because HBB prevents the C/O ratio from exceeding unity for all their models with a mass greater than 3.5~$\Msun$ rather than the models not being cool enough to form molecules. 
The models of \citet{Ventura10} have a very efficient HBB owing to their choice of convective model, the Full Spectrum of Turbulence prescription \citep[FST,][]{Canuto91} and little TDU due to how they treat the convective borders. 

As intermediate-mass AGB stars can experience both TDU and HBB, the complex interplay between the two processes can cause the star to become either oxygen rich or carbon rich. Stellar models calculated with different stellar evolution codes can display varied behaviours including differences in the efficiency of TDU and HBB which can alter the surface C/O ratio allowing for either carbon- or oxygen-rich compositions. This in turn means that the effect of the molecular opacities on the stellar evolution can vary greatly depending on model assumptions.  In contrast, low-mass AGB stars only experience TDU which serves to increase the surface C/O ratio. 

The aim of this paper is to expand upon the work of Ventura \& Marigo using intermediate-mass AGB models that show a different evolution of the surface C/O ratio. In particular, our low-metallicity models show very deep dredge up and become carbon rich during the AGB phase in contrast to the models of, for example, \citet{Ventura10}. Therefore, their conclusions are not necessarily applicable to our models. 
The paper is presented as follows. In Section 2 we describe the method of calculation we use  to model an AGB star and detail the molecular opacities used in the investigation. In Section 3 we present the results of the stellar models and in Section 4 we present the effect on the yields between the two opacity treatments. Section 5  we compare with previous studies and summarise our results.

\section{Numerical method}

A two step procedure is used to calculate each AGB stellar model. Firstly, we use the Mt Stromlo stellar evolutionary code \citep[][and references therein]{Karakas10} to calculate the stellar evolutionary sequences. Each stellar model was evolved from the zero-age main sequence to near the end of the TP-AGB phase when the majority of the convective envelope is lost by strong stellar winds. For convective regions we use the standard mixing length theory \citep{Bohm-Vitense58} with a mixing length parameter of $\alpha= 1.86$. The details of the procedure and the evolution code is described in \citet{Karakas10} with the following differences.  We update the low-temperature molecular opacities (described in detail in Section~\ref{sec:opacity}) and the high-temperature radiative opacity tables. To be consistent with the molecular opacity tables, we use OPAL radiative tables  \citep{Iglesias96} with a \citet{Lodders03} scaled-solar abundance. We do not include convective overshoot in the formal sense in order to obtain the third dredge-up. However, we follow the method described by \citet{Lattanzio86} and \citet{Frost96} to determine a neutral border to each convective boundary. In this method, the ratio of the temperature gradients is linearly extrapolated from the last two convective mesh points to the first radiative point. If the extrapolated value is greater than unity then the point is determined to be in the convective zone, if less than unity the point is considered to be in the radiative zone. This means that only one point can be added, per iteration, to the convective zone.

The calculated stellar evolutionary sequences are used as input into a post-processing nucleosynthesis code \citep[see][]{Cannon93,Lugaro04}. The nuclear network we use is based on the JINA Reaclib\footnote{https://groups.nscl.msu.edu/jina/reaclib/db/} database as of May 2012 \citep{Cyburt10}. It includes 589 reactions of 77 species from hydrogen to sulphur with a small group of iron-peak elements (Fe, Co and Ni).  An additional `species' $g$ is included in the network to account for the number of neutron captures occurring beyond $^{62}$Ni; this $g$ species simulates the $s$-process as a neutron sink. We use a scaled-solar initial composition from \citet{Asplund09} and assume a solar global metallicity of $Z_{\odot} = 0.015$ which is comparable to the solar global metallicity recommended by \citet{Asplund09} of $Z_{\odot} = 0.0142$. This differs from the \citet{Lodders03} solar metallicity of 0.01321 in the opacity tables as low-temperature opacity tables using the \citet{Asplund09} values are not available.
Throughout this paper, all abundance ratios are by number whereas individual species are in mass fraction. We assume the envelope abundances are equivalent to surface abundances. 

\subsection{Mass loss}

We assume the same mass-loss formula for all our models.
Mass loss prior to the AGB phase is included using the \citet{Reimers75} formula with $\eta_R = 0.4$. This leads to very little mass being lost before the AGB phase. For example, the intermediate-mass models presented here lose less than 1 per cent of their initial mass before they reach the early AGB phase. 

Mass loss is included during the AGB phase using the \citet{Vassiliadis93} mass-loss prescription where only their equation 2 is used. The empirical formula of \citet{Vassiliadis93} was determined using a sample of oxygen- and carbon-rich AGB stars in the Milky Way and Magellanic Clouds which cover a range of luminosities and pulsation periods.

\subsection{Molecular opacities}
\label{sec:opacity}

Bound-bound absorption by molecules becomes an important contribution to the total opacity at temperatures below 5000 K \citep{Alexander94}.
The abundance of carbon relative to oxygen in the stellar envelope can significantly influence the formation of molecules and has a substantial effect on the opacity at low temperatures. Changes in the molecular opacity occur when the stellar envelope transitions from an oxygen-rich (C/O $<$ 1) to a carbon-rich (C/O $>$ 1) chemical composition  \citep{Marigo09}. This is due to the excess of carbon atoms allowing for the formation of carbon-bearing molecules such as, for example, HCN, CN, C$_2$ and SiC. 

The intermediate-mass stellar models of \citet{Karakas12} use low-temperature molecular opacity tables from \citet{Lederer09} that account for only an increase in C due to TDU. For low-mass AGB stellar models that do not undergo HBB this method is sufficient as TDU will only increase the C abundance in the envelope. However, for intermediate-mass stars, CNO cycling during HBB causes a decrease in the C abundance (as well as the O abundance) in the envelope and it is possible that the C abundance will become lower than the initial C abundance. The subsequent decrease in O along with C can still cause the star to have a C/O ratio above unity. The opacity tables used in \citet{Karakas12} do not account for this. In order to account for these variations we include low-temperature opacity tables to follow the decrease in C and O due to HBB. 

For the stellar models presented here we use the \AE SOPUS low-temperature molecular opacity tables \citep{Marigo09} with a  \citet{Lodders03} scaled-solar abundance. The low-temperature opacity tables have been calculated to follow the variations in the chemical composition of C, N and O in the envelope due to TDU and HBB. The tables use Rosseland mean opacities and are a function of temperature log($T$) and log($R$) where $R=\rho/(T/10^6~{\rm  K})^3$ for an arbitrary chemical composition \citep{Marigo09}. We use linear interpolation between tables with log($T$)  from 3.2 to 4.05 in steps of 0.05 dex and log($R$) from $-$7.0 to 1.0 in steps of 0.5 dex.

To account for the changes in C, N and O in the envelope due to TDU and HBB, a variation factor $f_i$ is used,

\begin{equation}
X_i = f_iX_{i,{\rm ref}},
\end{equation}

where $X_i$ is the current abundance of species $i$ in mass fraction and $X_{i,{\rm ref}}$ is the initial reference abundance of species $i$. A value of $f_i > 1$ indicates an enhancement in the abundance whereas $f_i < 1$ indicates a depletion in the abundance compared to the initial reference abundance. The increase (or decrease) in abundance for C, N and O can be described using the following equations from \citet{Ventura09}: 

\begin{equation}
\left(\frac{X_C}{X_O}\right) = f_{C/O}\left(\frac{X_{C\rm{,ref}}}{X_{O\rm{,ref}}}\right),
\end{equation}

\begin{equation}
X_C = f_C~ X_{C\rm{,ref}},
\end{equation}

\begin{equation}
X_N = f_N~ X_{N\rm{,ref}}.
\end{equation}

The helium abundance is then given by $Y = 1 - X - Z$ so that the composition conserves mass where $X$ is the hydrogen abundance and $Z$ is the global metallicity. 

We perform tests using two opacity treatments: one which only accounts for the increase in C (referred to as $\kappa_{\rm C}$) and another where the increase and decrease in C and O are accounted for (referred to as $\kappa_{\rm CO}$). The $f_i$ values used in this study for C, N and C/O are detailed in Table~\ref{tab:opacity-tab}. We use the same scaled-solar  \AE SOPUS opacity tables as \citet{Kamath12} for the $\kappa_{\rm C}$ models. In the $\kappa_{\rm CO}$ models we incorporate two additional variation factors for C and the C/O ratio and explicitly follow the envelope C/O ratio to account for a possible reduction in C and O abundances due to HBB for the $\kappa_{\rm CO}$ models. 
The variations in abundance of N are treated the same for all models. 

\begin{table}
 \begin{center}
  \caption{Values of $f_i$ factors for the calculation of molecular opacity for the $\kappa_{\rm C}$ and $\kappa_{\rm CO}$ stellar models. The additional values included in the $\kappa_{\rm CO}$ treatment are shown in bold.
 \label{tab:opacity-tab}}
  \vspace{1mm}
   \begin{tabular}{llccccc}
\hline \hline
$Z$ & $X$ & log($f_{\rm C/O}$) & log($f_{\rm C}$)& log($f_{\rm N}$) & C/O \\ 
    &                                    &  &   &  \\
\hline
0.001 & 0.5 & {\bf -1.00}  & {\bf -1.00} & 0.00 & {\bf 0.050} \\ 
&0.6 & {\bf -0.35}  & {\bf -0.35} & 0.60 & {\bf 0.224} \\ 
&0.7 & 0.00  & 0.00 & 1.20 & 0.501 \\ 
&0.8 & 0.17 & 0.17 & & 0.741\\
 & &  0.30  & 0.30  &     &  1.000 \\
& &  0.35 & 0.35 & & 1.122  \\ 
 &&  0.90 & 0.90 &  & 3.980 \\ 
 &  &1.55 & 1.55 &  & 17.779 \\ 
 &  &2.17 & 2.17 & & 74.114 \\ 
 \\
0.02 & 0.5 & {\bf -1.00}  & {\bf -1.00} & 0.00 & {\bf 0.050} \\ 
&0.6 & {\bf -0.35}  & {\bf -0.35} & 0.60 & {\bf 0.224} \\ 
&0.7& 0.00  & 0.00 & 0.30 & 0.501 \\ 
&0.8 & 0.17 & 0.17 &  & 0.741\\
 &&  0.25  & 0.25  & &  0.891 \\
 &&  0.30 & 0.30 & & 1.000  \\ 
 &&  0.35 & 0.35 &  & 1.122 \\ 
 &&  0.50 & 0.50 &  & 1.585 \\ 
&& 0.70 & 0.70& & 2.511 \\ 

\hline \hline
  \end{tabular} 
\\
 \end{center}
\end{table}

\section{Stellar models} 

\subsection{Stellar structure} 

We model two stellar masses, namely 5~$\Msun$ and 6~$\Msun$, with two metallicities, $Z~=~0.001$ and $Z~=~0.02$, to determine the effect of the two different opacity treatments, $\kappa_{\rm C}$ and $\kappa_{\rm CO}$. 
All other input parameters in the stellar evolution code are kept to be the same so any differences in the stellar structure can be attributed to the different treatments of molecular opacity. There are negligible differences in the evolution of the stellar structure during the pre-AGB phase between the two opacity treatments and therefore only differences in the stellar models during the AGB phase will be discussed.
Relevant properties of the stellar structure calculations for each of the models are summarised in Table~\ref{tab:strucresults} and includes the number of TPs calculated, final core mass (M$_{\rm core}$), final envelope mass (M$_{\rm env}$), maximum temperature reached at the base of the convective envelope ($T_{\rm bce}^{\rm max}$) and the total amount of material dredged-up due to TDU (M$_{\rm dred}^{\rm tot}$). 

\begin{table}
 \begin{center}
  \caption{Properties of the calculated stellar models. Column 1 lists the opacity treatment, column 2 the initial mass, column 3 the initial global metallicity, column 4 the number of TPs calculated, column 5 the final core mass, column 6 the final envelope mass, column 7 the maximum temperature reached at the base of the convective envelope, and column 8 the total amount of material dredged up due to TDU.
 \label{tab:strucresults}}
\renewcommand{\tabcolsep}{1.1mm}
   \begin{tabular}{ccccccccc}
\hline\hline 
Opacity & Mass  & $Z$ &  TPs & Final M$_{\rm core}$& Final M$_{\rm env}$& $T_{\rm bce}^{\rm max}$ & M$_{\rm dred}^{\rm tot}$ \\ 
&   ($\Msun$)  &   &   &   ($\Msun$)  &($\Msun$)  & ($\times 10^6$ K) & ($\Msun$)  \\
\hline
$\kappa_{\rm C}$ & 5.0 &  0.001 & 108 &0.940 & 0.960 & 92.99 &0.216 \\ 
& 6.0 & 0.001 & 126 &1.018 & 0.888 & 104.91 & 0.127\\
& 5.0 &  0.02 & 30 &  0.868  & 0.713& 63.29 & 0.088 \\
& 6.0 & 0.02 & 41 & 0.907& 1.084 & 81.59 & 0.090\\ 
\\
$\kappa_{\rm CO}$ & 5.0 &  0.001 & 96 & 0.938 & 1.035 & 92.47 & 0.192 \\ 
& 6.0 & 0.001 & 112 &1.015 & 0.662& 104.79 & 0.107 \\
& 5.0 &  0.02 & 29  & 0.867  & 1.013 & 63.95  & 0.083\\
& 6.0 & 0.02 & 42  & 0.908  & 1.352 & 82.30 & 0.092 \\ 

\hline \hline
  \end{tabular} 
\\

 \end{center}
\end{table}

\subsubsection{$Z~=~0.001$ models}

\begin{figure}
\begin{center}
\includegraphics[width=\columnwidth]{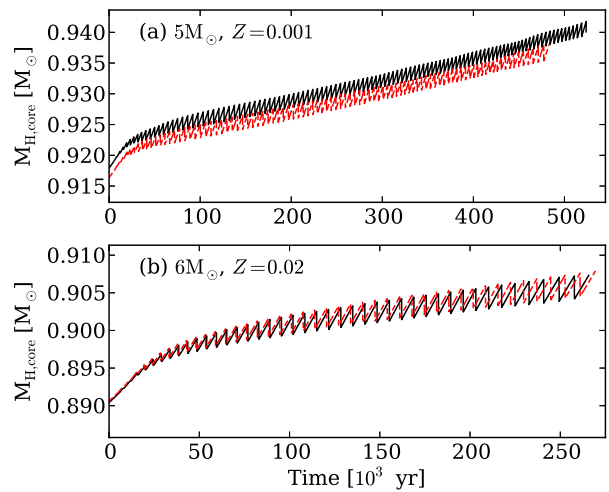} 
\caption{Evolution of the H-exhausted core (M$_{\rm H,core}$) from the start of the TP-AGB phase ($t=0$) for two of the stellar models calculated. The solid (black) lines show the $\kappa_{\rm C}$ models while the dashed (red) lines show the $\kappa_{\rm CO}$ models.}
\label{fig:Hexhaustedcore}
\end{center}
\end{figure}

 Figure~\ref{fig:Hexhaustedcore} (panel $a$) shows the evolution of the H-exhausted core with time for the 5~$\Msun$, $Z~=~0.001$ model. 
 The H-exhausted core mass at the beginning of the TP-AGB phase is slightly higher for the $\kappa_{\rm C}$ models, around 0.001~$\Msun$ for the 5~$\Msun$ model. This can be attributed to minor differences during core He-burning evolution.
  It can be seen that the updated opacity treatment $\kappa_{\rm CO}$ reduces the number of TPs during the AGB phase for the $Z~=~0.001$ models.
   The 5~$\Msun$, $Z~=~0.001$ $\kappa_{\rm CO}$ model has 12 fewer TPs while the 6~$\Msun$, $Z~=~0.001$ $\kappa_{\rm CO}$ model has 14 fewer TPs when compared to the $\kappa_{\rm C}$ models. This means that less enriched material is dredged up to the surface as demonstrated in Table~\ref{tab:strucresults} where the total amount of He-intershell material decreases by around 15 per cent for the $\kappa_{\rm CO}$ models. The $Z~=~0.001$ $\kappa_{\rm C}$ models each dredge up approximately 0.02~$\Msun$ more material over the lifetime of the AGB phase compared to the $\kappa_{\rm CO}$ models.

\begin{figure}
\begin{center}
\includegraphics[width=\columnwidth]{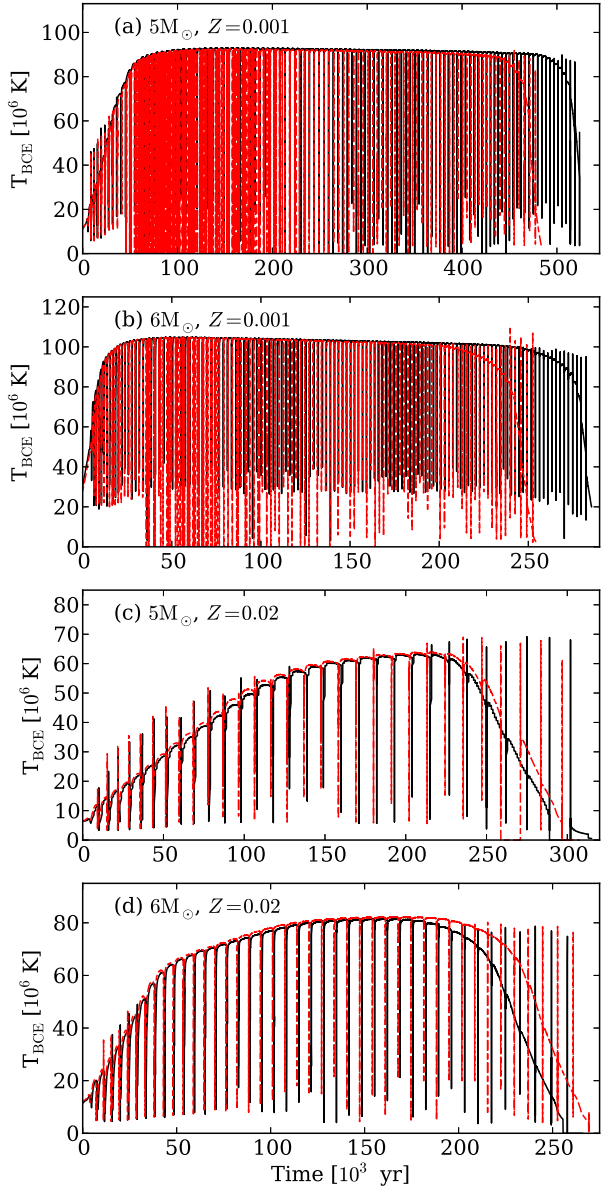} 
\caption{Evolution of the temperature at the base of the convective envelope from the start of the TP-AGB phase ($t=0$) for each stellar model calculated. The solid (black) lines show the $\kappa_{\rm C}$ models while the dashed (red) lines show the $\kappa_{\rm CO}$ models.}
\label{fig:TatBCE}
\end{center}
\end{figure}

The evolution of the temperature at the base of the convective envelope for each model is shown in Figure~\ref{fig:TatBCE} (panels $a$ and $b$). HBB is evident as the temperature reaches higher than the $(50 - 80) \times 10^6$ K required for CNO cycling at the base of the convective envelope.  Figure~\ref{fig:TatBCE} for the $Z~=~0.001$ models shows that, when using the updated opacity tables $\kappa_{\rm CO}$, HBB is extinguished earlier. This is because the $\kappa_{\rm CO}$ models become more extended in radius and therefore cooler. This change due to the increased molecular opacity leads to an enhanced mass-loss rate which ejects the envelope sooner.

\begin{figure}
\begin{center}
\includegraphics[width=\columnwidth]{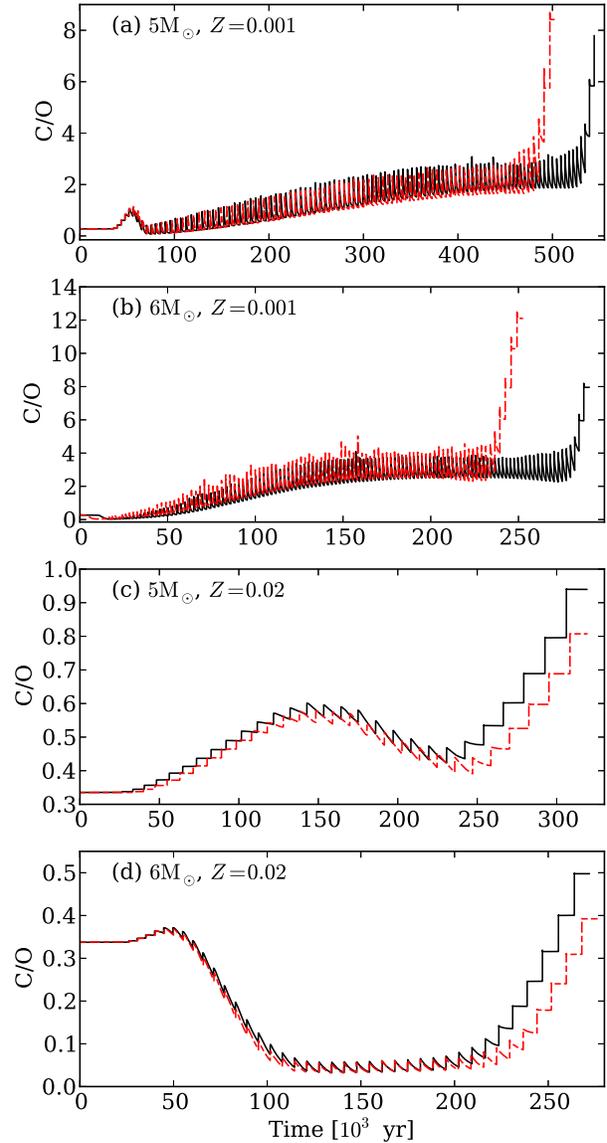} 
\caption{The variation in the C/O ratio with time for each of the stellar models calculated from the start of the TP-AGB phase ($t=0$). The solid (black) lines show the $\kappa_{\rm C}$ models while the dashed (red) lines show the $\kappa_{\rm CO}$ models.}
\label{fig:CO}
\end{center}
\end{figure}

\begin{figure}
\begin{center}
\includegraphics[width=\columnwidth]{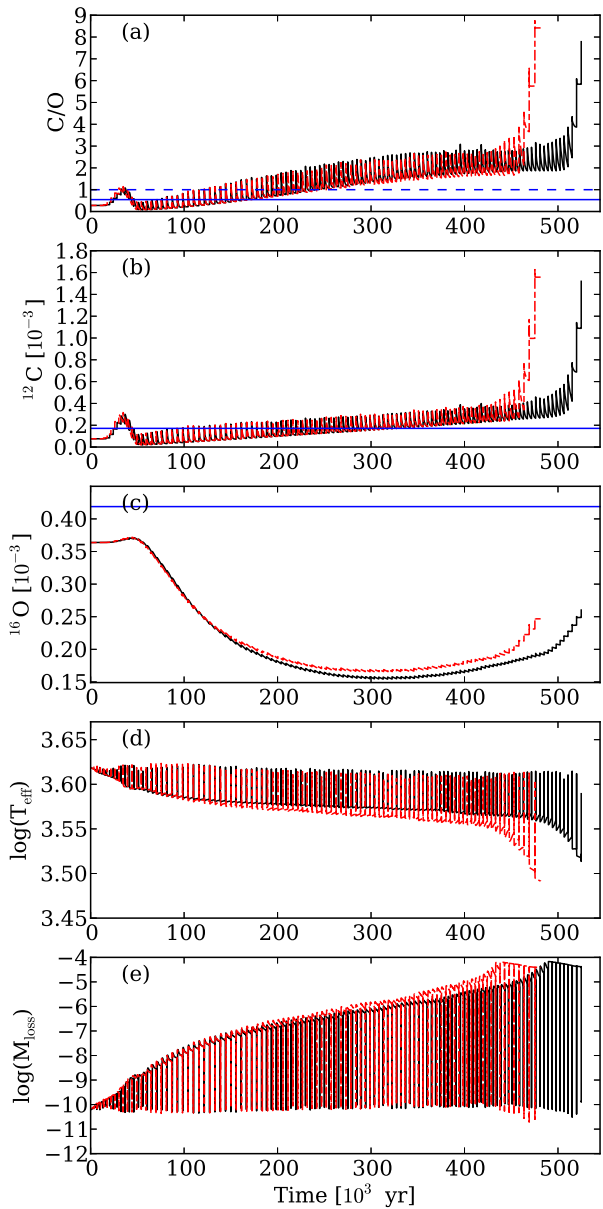} 
\caption{The 5~$\Msun$, $Z~=~0.001$ stellar model showing the variation in ($a$) C/O, ($b$) $^{12}$C surface abundance, ($c$) $^{16}$O surface abundance, ($d$) log$_{10}$(T$_{\rm eff}$) and ($e$) log$_{10}$(M$_{\rm loss}$)  with time from the start of the TP-AGB phase ($t=0$). The dashed (blue) line shows a C/O ratio of unity while the solid (blue) lines show the initial $^{12}$C and $^{16}$O surface abundance.}
\label{fig:CO-surf}
\end{center}
\end{figure}

Figure~\ref{fig:CO} (panels $a$ and $b$) compares the evolution of the C/O ratio between the two opacity treatments for both $Z~=~0.001$ models.
For a period during the AGB phase, the competing effects of TDU and HBB can cause the C/O ratio to fluctuate between below unity and above unity. For a brief time during this period the $^{12}$C surface abundance is below the initial $^{12}$C surface abundance while the C/O ratio is greater than unity. The molecular opacity is then underestimated in the $\kappa_{\rm C}$ model as the fact that the C/O ratio exceeds unity has not been taken into account during these conditions.  This omission in the determination of the opacity results in the discrepancies seen in the stellar evolution between the $\kappa_{\rm C}$ and $\kappa_{\rm CO}$ models in Figure~\ref{fig:CO-surf}.  Figure~\ref{fig:CO-surf} shows the temporal evolution of the C/O ratio, $^{12}$C surface abundance, $^{16}$O surface abundance, effective temperature and mass-loss rate for the 5~$\Msun$, $Z~=~0.001$ models.

As mentioned by \citet{Marigo02}, the consequences of using opacity tables that do not correctly follow the variation of the C/O ratio when the star is carbon rich include an inaccurate effective temperature. Figure~\ref{fig:CO-surf} shows that the difference between the $\kappa_{\rm C}$ and $\kappa_{\rm CO}$ models is not due to a slightly different core mass at the beginning of the TP-AGB phase. The lower effective temperature (panel $d$) and higher mass-loss rate (panel $e$) for the $\kappa_{\rm CO}$ models starts to become significant when the surface C/O ratio for the models changes from below unity to above. This is where the $\kappa_{\rm CO}$ molecular opacity treatment more realistically follows the increase in the C/O ratio.

\subsubsection{$Z~=~0.02$ models}

The effect of the different opacity treatment on the number of TPs is less pronounced for the $Z~=~0.02$ models than for the $Z~=~0.001$ models.  Figure~\ref{fig:Hexhaustedcore} (panel $b$) shows that the difference in the evolution of the mass of the hydrogen exhausted core between the two opacity treatments is very slight for the 6~$\Msun$ model. The differences between the two opacity treatments are minimal and the C/O ratio of the models does not exceed unity which would result in a higher molecular opacity where carbon-bearing molecules dominate.  Compared to the $Z~=~0.001$ models, there are fewer TPs with a shorter lifetime of the TP-AGB phase. The 5~$\Msun$ $\kappa_{\rm CO}$ model experiences one less TP while the 6~$\Msun$ $\kappa_{\rm CO}$ model experiences one more TP than the $\kappa_{\rm C}$ models. We conclude that the differences in the number of TPs is insignificant. 

\begin{figure}[h]
\begin{center}
\includegraphics[width=\columnwidth]{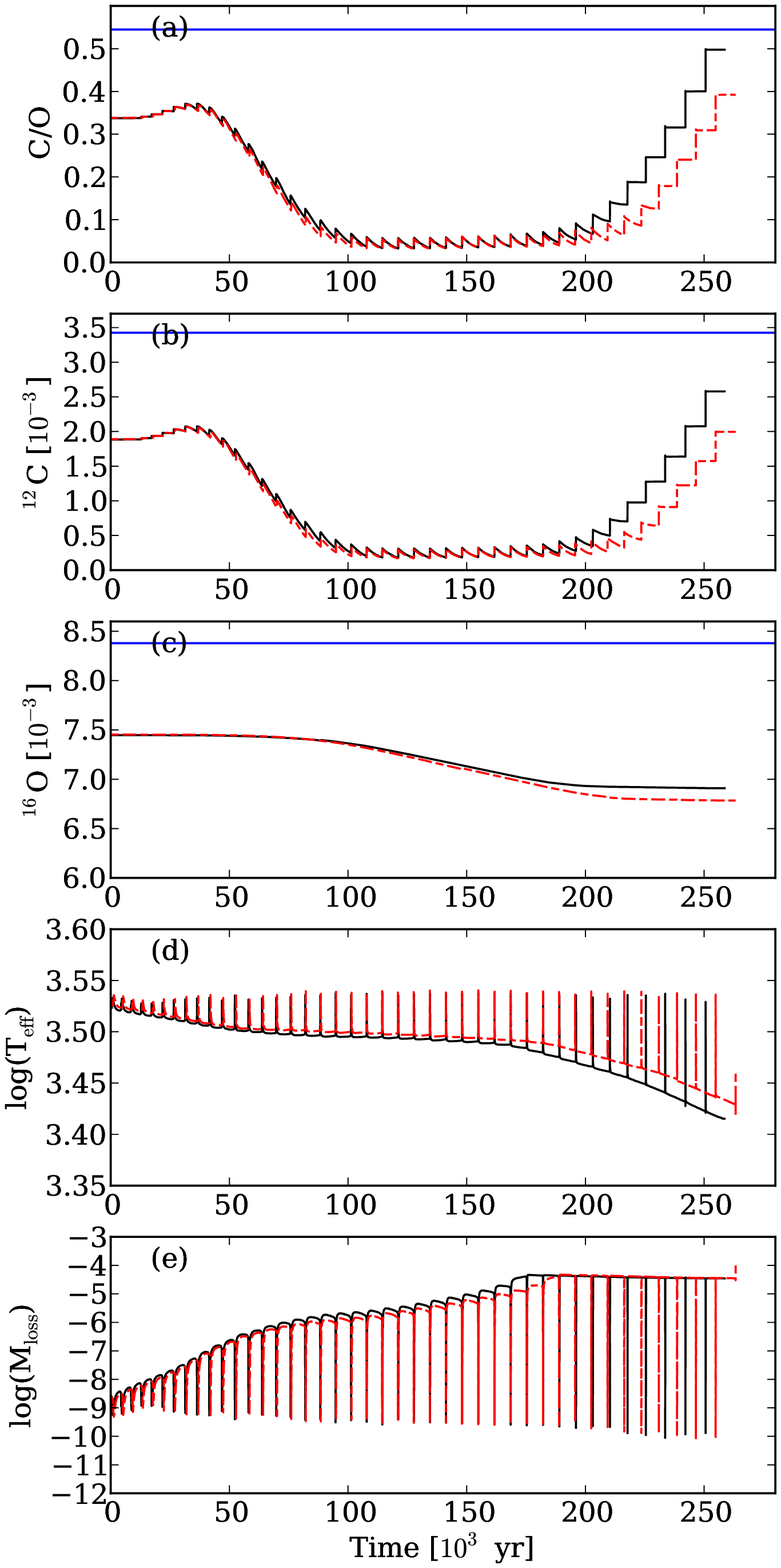} 
\caption{The 6~$\Msun$, $Z~=~0.02$ stellar model showing the variation in ($a$) C/O, ($b$) $^{12}$C surface abundance, ($c$) $^{16}$O surface abundance, ($d$) log$_{10}$(T$_{\rm eff}$) and ($e$) log$_{10}$(M$_{\rm loss}$)  with time from the start of the TP-AGB phase ($t=0$). The solid (blue) lines show the initial $^{12}$C and $^{16}$O surface abundance.}
\label{fig:co_m6z02}
\end{center}
\end{figure}

Figure~\ref{fig:co_m6z02} shows the evolution of the C/O ratio, $^{12}$C surface abundance, $^{16}$O surface abundance, effective temperature and mass-loss rate for the TP-AGB phase for the 6~$\Msun$ model. Pre-AGB evolution mixing episodes cause the surface $^{12}$C and $^{16}$O abundances to decrease below initial values. This causes the C/O ratio to decrease with the C/O ratio being below the initial value at the start of the TP-AGB phase. 
HBB is only active for part of the TP-AGB evolution in the 6~$\Msun$, $Z~=~0.02$  models. This is illustrated in Figure~\ref{fig:co_m6z02} (panel $a$) where the C/O ratio initially increases, albeit with a shallow gradient and then decreases once the temperature at the base of the envelope becomes hot enough for proton captures onto $^{12}$C. Eventually HBB is extinguished and the C/O ratio increases to almost its initial value owing to the continuation of TDU. It is when HBB is active that the $\kappa_{\rm C}$ model does not accurately follow the variation of the C and O abundance as only the increase in the $^{12}$C abundance is taken into account.

\begin{figure}[h]
\begin{center}
\includegraphics[width=\columnwidth]{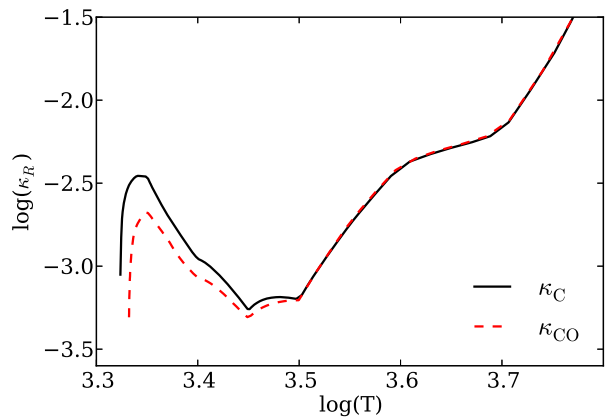} 
\caption{The difference in opacity $\kappa_R$ between the 6~$\Msun$, $Z~=~0.02$ $\kappa_{\rm C}$ (solid black line) and $\kappa_{\rm CO}$ (dashed red line) models for the same total mass (approximately~2.3~$\Msun$). The opacity is noticeably different for log($T$) $\lesssim$ 3.5 near the stellar surface between the two opacity treatments. }
\label{fig:opacity-m6z02}
\end{center}
\end{figure}

 For the $\kappa_{\rm C}$ 6~$\Msun$ model this situation where the current C/O, C and O abundances are never greater than the initial abundances is treated as if the surface abundance has a scaled-solar composition. The low-temperature molecular opacity does not vary for a given temperature and density with the variation factors being $f_C = f_O = 0$.  Figure~\ref{fig:opacity-m6z02} shows the opacity below log($T$) $\lesssim$ 4 in the 6~$\Msun$ $\kappa_{\rm C}$ and $\kappa_{\rm CO}$ models for the same total stellar mass of approximately~2.3~$\Msun$. This snapshot was chosen where the opacity difference is largest. At log($T$) $\lesssim$~3.5 near the stellar surface, the opacity is lower for the $\kappa_{\rm CO}$ model compared to the $\kappa_{\rm C}$ model. This decrease in opacity leads to a higher effective temperature (as seen in the bottom panel of Figure~\ref{fig:co_m6z02}) and as a consequence the mass-loss rate is lower and HBB ceases slightly later. The sustained level of HBB causes a lower final C/O ratio compared to the $\kappa_{\rm C}$ models as seen in Figure~\ref{fig:co_m6z02} (panel $a$). 

\subsection{Stellar yields}

\begin{table*}
\small
 \begin{center}
  \caption{Net yields of selected isotopes. Yields are in solar masses and are expressed in the form $n(m) = n \times 10^m$. 
 \label{tab:yieldresults}}
  \vspace{1mm}
  
   \begin{tabular}{cccccccccccc}
\hline \hline
Opacities & Mass  & $Z$ &  $^{12}$C & $^{13}$C & $^{14}$N & $^{16}$O & $^{22}$Ne & $^{23}$Na   & $^{25}$Mg & $^{26}$Mg & $g$\\ 
\hline 
$\kappa_{\rm C}$ & 5.0 &  0.001 & 1.652(-3)	& 5.718(-4)	& 5.715(-2)	& -8.273(-4)	& 1.845(-3)	& 4.725(-5)	&  3.399(-4)	& 2.866(-4) & 1.793(-6) \\ 
& 6.0 & 0.001 & 9.784(-4)	& 4.298(-4)	& 3.806(-2)	& -1.547(-3)	& 5.404(-4)	& 6.487(-6)	    & 2.197(-4)	& 1.279(-4)	&   1.241(-6)  \\
& 5.0 &  0.02 &  1.864(-3)	& 3.476(-3)	& 1.971(-2)	& -2.831(-3)	& 1.720(-3)	& 9.558(-5)  	& 1.563(-4)	& 8.990(-5)	& 8.321(-7) \\
& 6.0 & 0.02 & -1.112(-2)	& 5.849(-4)	& 4.305(-2)	& -7.224(-3)	& 1.637(-3)	& 1.350(-4)	& 2.039(-4)	& 1.248(-4)	& 1.268(-6) \\ 
\\
$\kappa_{\rm CO}$ & 5.0 &  0.001 &  2.101(-3)	& 5.447(-4)	& 5.030(-2)	& -8.422(-4)	& 1.495(-3)	& 3.763(-5)		& 2.991(-4)	& 2.292(-4)& 1.473(-6) \\ 
& 6.0 & 0.001 &  1.309(-3) 	& 3.877(-4)	& 3.091(-2)	& -1.621(-3)	& 4.166(-4)	& 2.717(-6)		& 1.989(-4)	& 8.397(-5)	& 8.665(-7) \\
& 5.0 &  0.02 & 3.780(-4) 	& 3.238(-3)	& 2.156(-2)	& -2.862(-3)	& 1.691(-3)	& 9.566(-5)		& 1.512(-4)	& 8.759(-5)	& 8.183(-7)\\
& 6.0 & 0.02 &  -1.207(-2)  	& 5.589(-4)	& 4.509(-2)	& -7.817(-3)	& 1.665(-3)	& 1.332(-4)	& 2.281(-4)	& 1.239(-4)	& 1.304(-6) \\ 

\hline \hline
  \end{tabular} 
\\
 \end{center}
\end{table*}

In order to determine the total amount of material ejected into the interstellar medium during the lifetime of the stellar model, we calculate the net yield $M_i$ (in solar masses) of species $i$ to be,

\begin{equation}
M_i = \int_0^{\tau} \left [ X(i) - X_0(i) \right ] \frac{dM}{dt} dt,
\end{equation}

where $dM/dt$ is the current mass-loss rate in $\Msun$~yr$^{-1}$, $X(i)$ and $X_0(i)$ are the current and initial mass fraction of species $i$, and $\tau$ is the total lifetime of the stellar model in years. After calculating the stellar yields we calculate the percentage change $\Delta M_i$ for each stable isotope where $M_{i,{\rm C}}$ is the yield for species $i$ using the $\kappa_{\rm C}$ models and $M_{i,{\rm CO}}$ is the yield for species $i$ using the $\kappa_{\rm CO}$ models. The percentage change in yields between the $\kappa_{\rm C}$ and $\kappa_{\rm CO}$ models for each mass and metallicity is shown in Figure~\ref{fig:yields}.  The changes in surface abundance, and consequently the final net yield, are dependent on the amount of material dredged up from the He-intershell as well as the maximum temperature and duration of HBB. Yields for selected isotopes are presented in Table~\ref{tab:yieldresults} for the two opacity treatments, $\kappa_{\rm C}$ and $\kappa_{\rm CO}$.

\subsubsection{$Z~=~0.001$ yields}

Figure~\ref{fig:yields} (panels $a$ and $b$) shows that the 5~$\Msun$ and 6~$\Msun$, $Z~=~0.001$ models display a similar trend in yield differences between the two opacity treatments, $\kappa_{\rm C}$ and $\kappa_{\rm CO}$. Percentage changes of up to 30 per cent are seen for the 5~$\Msun$ model whereas the 6~$\Msun$ model shows percentage changes of up to 60 per cent in the yield. However the yield of $^{19}$F decreases by 250 per cent for the 5~$\Msun$ models which is not seen in the 6~$\Msun$ models.
 The reduced number of TDU episodes for the $\kappa_{\rm CO}$ influences the changes seen in the yields.  A higher percentage change is seen in the yields for the 6~$\Msun$ model compared to the 5~$\Msun$ model. This is caused by a larger decrease in the number of TPs as well as the shorter duration of HBB between the $\kappa_{\rm C}$ and $\kappa_{\rm CO}$ models.  
 
 The isotopes $^{12}$C and $^{16}$O are produced in the intershell region through partial He-burning and are brought to the surface by repeated TDU. During the interpulse period, HBB produces $^{14}$N at the expense of $^{12}$C and $^{16}$O. 
 Due to a shorter period of HBB for the $\kappa_{\rm CO}$ models, the $^{12}$C yield is higher despite these models dredging up less material (see Table~\ref{tab:strucresults}). 
  Therefore the $^{14}$N yield is also lower for the $\kappa_{\rm CO}$ models for the same reason. The majority of $^{13}$C is produced during HBB through the CN cycle before being destroyed via the $^{13}$C(p,$\gamma$)$^{14}$N reaction. The $^{13}$C yield decreases for the $\kappa_{\rm CO}$ models due to a shorter period of HBB combined with the slightly lower temperature at the base of the convective envelope.  Even though $^{16}$O is being dredged up to the surface the amount is much less than $^{12}$C as the intershell composition comprises of $\lesssim$ 1 per cent $^{16}$O and around 25 per cent $^{12}$C.  The majority of the $^{16}$O is destroyed through HBB and the net yield of $^{16}$O is negative for both opacity treatments. 

The isotope $^{22}$Ne is produced in the intershell via $\alpha$-captures onto $^{14}$N. The amount of $^{22}$Ne mixed into the envelope affects the yields of isotopes in the Ne-Na chain. For example, $^{23}$Na is formed by proton-captures onto $^{22}$Ne. As the $\kappa_{\rm CO}$ models experience fewer TDU episodes, the yields of $^{22}$Ne and $^{23}$Na are lower than for the $\kappa_{\rm C}$ models by around 20 per cent each for the 5~$\Msun$ model. For the 6~$\Msun$ $\kappa_{\rm CO}$ model the yield of $^{22}$Ne is lower by around 20 per cent and for $^{23}$Na, the yield decreases by around 60 per cent as seen in Figure~\ref{fig:yields}. The isotopes $^{25}$Mg and  $^{26}$Mg  are produced through the $^{22}$Ne($\alpha$,n)$^{25}$Mg and $^{22}$Ne($\alpha$,$\gamma$)$^{26}$Mg reactions which are activated at temperatures above $3 \times 10^8$ K. These conditions occur in the convective region that develops during a TP. 
Less synthesised $^{25}$Mg and $^{26}$Mg reaches the surface for the $\kappa_{\rm CO}$ models resulting in lower yields with an approximately 10 to 30 per cent decrease compared to the $\kappa_{\rm C}$ models. 

Any change in the abundance of the iron group isotopes is due to neutron-capture through the $s$-process. The isotope $^{56}$Fe is used as a seed for the $s$-process during a convective TP where neutrons are released by the $^{22}$Ne($\alpha$,n)$^{25}$Mg reaction. Any neutron captures onto $^{56}$Fe or isotopes heavier than  $^{56}$Fe are accounted for by the neutron sink $g$. Therefore $g$ can be thought of as the sum of abundances for the $s$-process isotopes. 
For the $\kappa_{\rm CO}$ models the yield of the neutron sink $g$ is lower for the $\kappa_{\rm CO}$ models compared to the $\kappa_{\rm C}$ models by  18 per cent for the 5~$\Msun$ model and 30 per cent for the 6~$\Msun$ model. These yield changes can all be attributed to the reduced number of TDU episodes.

\subsubsection{$Z~=~0.02$ yields}

Figure~\ref{fig:yields}  (panels \emph{c} and \emph{d}) shows yield differences of up to 20 per cent for lighter isotopes with an atomic mass up to about 18 including the CNO isotopes  for the $Z~=~0.02$ models..  For $^{19}$F we find a difference of around 15 per cent for the 6~$\Msun$ model with a smaller difference for the 5~$\Msun$ model.  For the intermediate-mass isotopes, which include here $^{22}$Ne, $^{23}$Na, $^{25}$M, $^{26}$Mg and $^{26}$Al$^g$ we find very small yield differences of less than 10 per cent for both the 5~$\Msun$ and 6~$\Msun$ models as illustrated in Figure~\ref{fig:yields}. 
There is a negligible effect on the net yield of the neutron sink $g$ with a percentage change of less than 5 per cent.

From Figure~\ref{fig:yields} we can see that the relative size of the yield changes between the $\kappa_{\rm C}$ and $\kappa_{\rm CO}$ models are smaller than for the $Z~=~0.001$ models. This is mainly because the $\kappa_{\rm C}$ and $\kappa_{\rm CO}$ $Z~=~0.02$ models experience almost the same number of TPs (as shown in Table~\ref{tab:strucresults}). In Table~\ref{tab:yieldresults} we see that the $^{12}$C yields of the 5~$\Msun$ models are positive compared to the negative $^{12}$C yields of the 6~$\Msun$ models. This indicates that there is an overall net production of carbon in the 5~$\Msun$ models. 
 While the amount of material dredged into the envelope is similar in both the 5~$\Msun$ and 6~$\Msun$ models (e.g., see Table~\ref{tab:strucresults}), the maximum temperature at the base of the envelope in the 5~$\Msun$ is considerably lower. This means that HBB is not very efficient in the 5~$\Msun$ as shown in Figure~\ref{fig:TatBCE}, where the maximum temperature is only $64 \times 10^6$ K compared to $82 \times 10^6$ K for the 6~$\Msun$ models and this temperature is only sustained for about 5 TPs. This means that the carbon dredged to the surface is not as efficiently burnt via the CNO cycles  at the base of the envelope.

However, both $\kappa_{\rm CO}$ models have a decrease in the $^{12}$C yields  compared to the $\kappa_{\rm C}$ models. In particular, the 5~$\Msun$ $\kappa_{\rm CO}$ model has a decrease of 80 per cent compared to the $\kappa_{\rm C}$ model. This decrease in $^{12}$C can be attributed to the longer duration, as well the higher temperatures, of HBB. This is despite the slight differences in the number of TDU episodes.  The negative yield of $^{16}$O indicates that more is being destroyed through HBB than being mixed from the intershell region to the surface. The longer duration of HBB also explains the increase in the $^{14}$N yield as more $^{12}$C is burnt to produce $^{14}$N via the CNO cycles.

\section{Discussion and Conclusions}

 The demand for accurate AGB models requires a more thorough understanding of the uncertainties in the input physics which play a role in determining the stellar evolution of AGB stars. Here we present new detailed evolutionary models of two masses, 5~$\Msun$ and 6~$\Msun$, and two metallicities, $Z~=~0.001$ and $Z~=~0.02$. These models are used to investigate the uncertainties in the stellar evolution when using two differing molecular opacity treatments, $\kappa_{\rm C}$ and $\kappa_{\rm CO}$, as well as investigating the effects of differing molecular opacity on stellar mass and metallicity. We determine that the AGB lifetime for the lower metallicity models is affected by the choice of the molecular opacity prescription.  The increased molecular opacity has the effect of lowering the effective temperature in the low-metallicity AGB models which, in turn, has the consequence of increasing the mass-loss rate. 
This consequence is confirmed by \citet{Marigo02} where the increase in the mass-loss rate due to increased molecular opacity serves to better explain the presence of carbon-rich stars in a population of Galactic giant stars. 
  In addition, previous studies by \citet{Ventura09} and \citet{Ventura10} highlight the importance of using variable abundance molecular opacities only if the C/O ratio exceeds unity. For low-mass AGB models where HBB does not occur it is sufficient for the molecular opacity to only follow the increase in carbon due to TDU as with the $\kappa_{\rm C}$ opacity tables. 

Previous studies, such as the stellar models by  \citet{Karakas12}, calculate the evolution of intermediate-mass AGB stars using the Mt Stromlo stellar evolutionary code but utilise opacity tables that only account for an increase in carbon due to TDU. 
This situation has been modelled here using the $\kappa_{\rm C}$ opacity tables. 
For the case when the surface C/O ratio does not exceed the initial C/O ratio, the $\kappa_{\rm C}$ opacity is calculated as if scaled-solar opacity tables had been used. The $\kappa_{\rm CO}$ opacity tables are able to interpolate between lower values  and as a consequence of the decreased opacity the stellar structure is affected. 
 The 6~$\Msun$, $Z~=~0.02$ model suffers from this omission of lower values below the scaled-solar composition in the low-temperature molecular opacity tables.  
However, as mentioned by \citet{Marigo07}, the evolution depends on the sensitivity of the mass-loss prescription to the effective temperature.

The studies by \citet{Stancliffe04} and Pignatari et al. (2013, submitted) calculate $Z~=~0.02$ intermediate-mass AGB models which experience efficient TDU. Our study finds that with efficient TDU the surface C/O ratio for the low-metallicity models can exceed unity even with the presence of efficient HBB. 
The study by \citet{Herwig04a} of $Z~=~0.0001$ intermediate-mass AGB models also finds efficient TDU where the C/O ratio exceeds unity for periods of time despite efficient HBB. This is in contrast to \citet{Ventura10} who find a threshold at $M~\geq~3.5~\Msun$ for a metallicity of $Z~=~0.001$ where, above this mass limit, the final C/O ratio of the models does not exceed unity. However, the Ventura \& Marigo models use an $\alpha$-enhanced composition and this could affect the ability of the models to become carbon rich. 

\begin{figure*}
\begin{center}
\includegraphics[width=2\columnwidth]{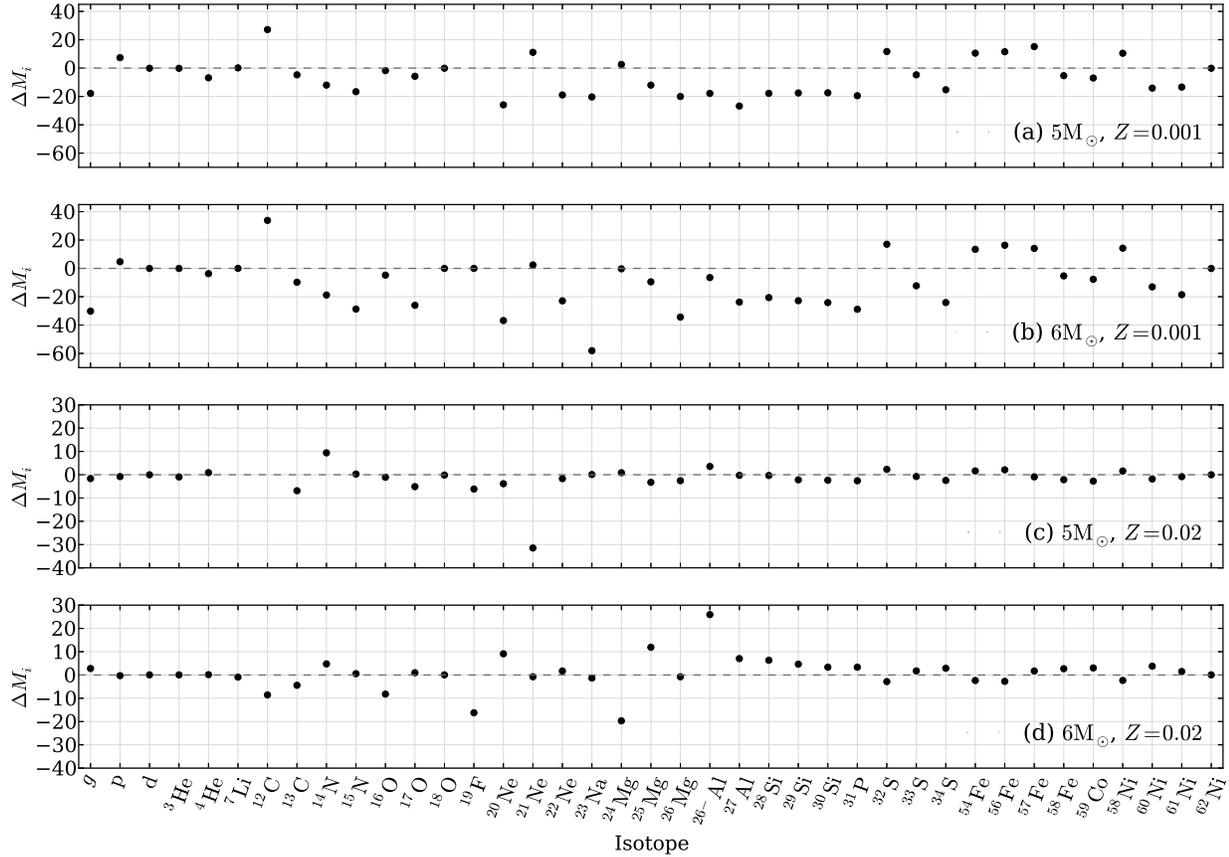} 
\caption{The percentage change in yield for species $i$ between the $\kappa_{\rm C}$ and $\kappa_{\rm CO}$ models for each model calculated where $\Delta M_i = 100(M_{i,{\rm CO}} - M_{i,{\rm C}})/|M_{i,{\rm C}}|$.  For the 5~$\Msun$ $Z~=~0.001$ model the percentage change for $^{19}$F is not shown. For the $\kappa_{\rm CO}$ model, the $^{19}$F yield decreases by around 250 per cent. For the 5~$\Msun$, $Z~=~0.02$ models the percentage change in $^7$Li and $^{12}$C is not shown. For the $\kappa_{\rm CO}$ model the yield of $^{7}$Li decreases by around 100 per cent while the yield of $^{12}$C decreases by around 80 per cent compared to the $\kappa_{\rm C}$ model. Each plot has the same $y$-axis scale for each metallicity for ease of comparison.}
\label{fig:yields}
\end{center}
\end{figure*}

We also investigated the effect of the two different molecular opacity treatments on the stellar yield predictions for 77 species. As illustrated in Figure~\ref{fig:yields} the net yields of the low-metallicity models are affected by changes in the stellar evolution due to opacity. The changes in the net yields of the $Z~=~0.02$ models are negligible when using the updated $\kappa_{\rm CO}$ opacity treatment. This result indicates the possibility that the changes in yield could be more considerable at a lower metallicity than $Z~=~0.001$.
 The $Z~=~0.001$ models of \citet{Marigo07} utilise synthetic TP-AGB models to investigate the effects of molecular opacity tables. \citet{Marigo07} concludes that the use of variable molecular opacities should not considerably affect the stellar yields in massive AGB models with $Z \leq 0.001$ as effective temperatures may be so high that the molecules cannot form. We do not find this to be the case for our $Z~=~0.001$ models as the effective temperatures are comparable to low-mass AGB models at the same metallicity.
 
  The treatment of molecular opacity along with other input physics such as the choice of the mass-loss rate, reaction rates and treatment of convection all contribute to the uncertainty in AGB models. These uncertainties can have a larger effect at $Z~=~0.02$ when compared to the opacity treatment. \citet{Ventura05a} investigate the effect of the treatment of convection on a 5~$\Msun$, $Z~=~0.001$ model. They find that the yields strongly depend on the convection model used with significant yield differences between MLT and FST. A previous study on the choice of mass-loss prescription by \citet{Stancliffe07} finds yield changes of  around 15 to 80  per cent for the light elements when investigating the effect on the yields for a 1.5~$\Msun$, $Z~=~0.008$ model. These differences are comparable to the results presented in this paper for the $Z~=~0.001$ models. The investigation on mass loss by \citet{Karakas06} finds yield changes of up to 90 per cent for a 5~$\Msun$, $Z~=~0.0001$ model. When looking at uncertainties in reaction rates for the Ne-Na and Mg-Al chains, \citet{Izzard07} find variations of up to two orders of magnitude for some species in the synthetic models. The models of \citet{Karakas06} find maximum yield differences up to 350 per cent when looking at different $^{22}$Ne + $\alpha$ reaction rates for a 5~$\Msun$, $Z~=~0.02$ model using the \citet{Reimers75} mass-loss prescription which is different to that used in this study. The yield differences due to reaction rate uncertainties are significantly higher when compared to uncertainties  found here due to the treatment of molecular opacities.  \citet{Marigo13} investigate a number of uncertainties including reaction rates, suppression of TDU and increasing the C and O abundance in the He-intershell for a 5~$\Msun$, $Z~=~0.001$ model. All the models tested become carbon rich and the evolution code is able to accurately follow the surface abundances of C, N and O as the molecular opacity is determined using `on-the-fly' calculations rather than through the interpolation of tables. The evolution of the surface abundance of a number of light elements as well as the temperature at the base of the convective envelope is shown to be dependent on the tested input parameters.
 However, with all these sources of inaccuracies, mass loss and convection are thought to dominate the uncertainties in stellar modelling. 

One use of stellar models is to provide theoretical predictions of chemical yields for comparison to observational data. Chemical yields are incorporated into chemical evolution models in order to understand the contribution of stellar populations to, for example, the Galaxy or globular clusters \citep{Kobayashi11,Cescutti12}. We have shown that the yields of intermediate-mass AGB stars can be affected by the treatment of molecular opacity in quite a complex way with changes of up to 20 per cent for most isotopes. These results also show that the degree of the difference depends on mass and metallicity. 
 Therefore it is more suitable to update current stellar models that experience HBB and become carbon rich to include molecular opacity tables that account for the changes in the surface abundances of C and O, as well as N, due to the CNO cycle that occurs during HBB.

\section*{Acknowledgments}

The authors are grateful to the referee for their helpful comments and careful reading of the paper. CKF thanks Anibal Garc\'{i}a-Hern\'{a}ndez for his useful comments and suggestions. 
AIK is grateful for the support of the NCI National Facility at the ANU and thanks the ARC for support through a Future Fellowship (FT10100475). RJS is the recipient of a Sofja Kovalevskaja Award from the Alexander von Humboldt Foundation. This research has made use of NASA's Astrophysics Data System.

\bsp

\label{lastpage}

\end{document}